\documentclass[doublespacing]{elsartMod}

\usepackage{graphicx}
\usepackage{amssymb}

\def\abs#1{\vert #1 \vert}

\begin{document}
\begin{frontmatter}


\journal{Physica {\bf B} (SCES'2001 proceedings)}


\title{Field-induced order and magnetization plateaux in frustrated
       antiferromagnets}

%
%
%
%
%
%

\author[DE]{A.\ Honecker\corauthref{1}\thanksref{DFG}}
\author[GB]{O.A.\ Petrenko}
\author[FR]{M.E.\ Zhitomirsky}

%
 
\address[DE]{TU Braunschweig, Institut f\"ur Theoretische Physik,
         Germany}
\address[GB]{University of Warwick, Department of Physics, Coventry, 
         UK}
\address[FR]{European Synchrotron Radiation Facility, 
         Grenoble, France}

%
%
%
%

\thanks[DFG]{Participation in SCES2001 supported by
the DFG.} 

%
%
%
%

\corauth[1]{Email: a.honecker@tu-bs.de}


\begin{abstract}

We argue that collinearly ordered states which exist in strongly frustrated
spin systems for special rational values of the magnetization are stabilized
by thermal as well as quantum fluctuations. These general predictions
are tested by Monte Carlo simulations for the classical and Lanczos
diagonalization for the $S={1 \over 2}$ frustrated square-lattice
antiferromagnet.

\end{abstract}

%
%

\begin{keyword}

field-induced order \sep magnetization plateaux

\end{keyword}


\end{frontmatter}

%
%
%
%
%

Frustrated 2D quantum spin systems can give rise to interesting
behavior in high magnetic fields. For example, plateaux in
the magnetization curve of SrCu$_2$(BO$_3$)$_2$
have been observed at\footnote{We normalize the magnetization to
$\abs{\langle M \rangle} \le 1$.}
$\langle M \rangle = {1 \over 8}$, ${1 \over 4}$ and ${1 \over 3}$
\cite{Kageyama-Onizuka}.
This has sparked theoretical interest in the magnetization process of the
related 2D Shastry-Sutherland model (see e.g.\ \cite{MoTo}).

Owing to its simplicity, the frustrated square-lattice antiferromagnet (FSAFM)
in a magnetic field is a useful model system \cite{prl,proc}:
\begin{equation}
H = J \sum_{\langle \vec{x}, \vec{y} \rangle}
 \vec{S}_{\vec{x}} \cdot \vec{S}_{\vec{y}}
+ J' \sum_{[ \vec{x}, \vec{y} ]}
 \vec{S}_{\vec{x}} \cdot \vec{S}_{\vec{y}}
 - h \sum_{\vec{x}} S^z_{\vec{x}} \, .
\label{H}
\end{equation}
Here the $\vec{S}_{\vec{x}}$ are spin-$S$ operators at the
sites $\vec{x}$ of a square lattice and the first and second sums are
over nearest and diagonal neighbor pairs, respectively. This model 
is realized with $S={1 \over 2}$ in Li$_2$VOSiO$_4$ and Li$_2$VOGeO$_4$
\cite{MCLMTMM}.

For $J' = {J \over 2}$ the Hamiltonian (\ref{H}) can be rewritten (up to a
constant) as
\begin{equation}
H = {1 \over 2^n} \sum_\alpha \left\{ J \vec{L}_\alpha^2 - h L_\alpha^z
 \right\}
\label{Hplaq}
\end{equation}
with $\vec{L}_\alpha = \sum_{\vec{x} \in \alpha} \vec{S}_{\vec{x}}$ the total
spin of a (four-spin) plaquette $\alpha$ and $n=2$.
In two dimensions the Kagome and checkerboard lattice
and in three dimensions the pyrochlore and garnet lattices share this
special form of the Hamiltonian.

Consider first the classical variant of the model (\ref{Hplaq}) with unit spins.
Then, any state with $L_\alpha^z = {h \over 2J}$, $L_\alpha^x = L_\alpha^y = 0$
for all $\alpha$ is a groundstate, resulting in huge degeneracies. 
Special collinear groundstates exist at special values of
the magnetic field (an up-up-up-down `uuud' state for
$\langle M \rangle={1 \over 2}$
in the FSAFM, checkerboard and pyrochlore lattice and an up-up-down `uud' state
for $\langle M \rangle={1 \over 3}$ in the Kagome and garnet lattice).
These collinear states have a particularly large number of zero modes leading
to an exceptionally large number of low-energy excitations above them.
At finite temperature, the probability function is therefore
peaked close to the collinear states along the groundstate manifold
leading to a selection of this submanifold by thermal fluctuations.
The subsequent lifting of degeneracy goes differently
for the FSAFM and Kagome AFM. In the former case a unique
classical state is further selected \cite{prl}, whereas in the latter
case there is no such selection and a collinear spin-liquid state
appears on the $\langle M \rangle={1 \over 3}$ plateau \cite{mzh}.

Semiclassical analysis \cite{prl} suggests that collinear states are also
stabilized by quantum fluctuations, leading to a plateau in the
magnetization curve of the quantum spin model. In fact, plateaux with
$\langle M \rangle={1 \over 2}$ are expected in the quantum FSAFM,
checkerboard and pyrochlore lattice and with $\langle M \rangle={1 \over 3}$
in the Kagome and garnet lattice even if they may not correspond to
collinearly ordered states. An $\langle M \rangle={1 \over 3}$ plateau is
indeed observed in exact diagonalization studies of the $S={1 \over 2}$ Kagome
lattice \cite{Wal-privcom}.

\begin{figure}
\centering
\includegraphics[width=7.5 true cm]{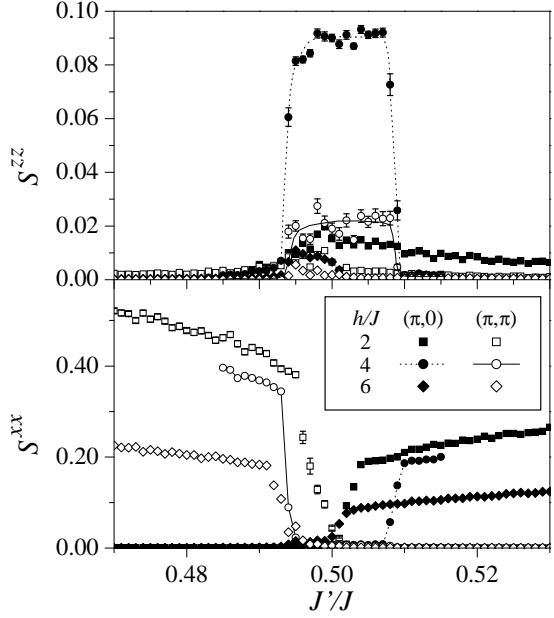}
\caption{Longitudinal (top panel) and transverse (bottom panel)
components of the static structure factor $S^{\alpha\beta}(\vec{q})$
for the classical FSAFM at $T=0.1J$.}
\label{SCfig}
\end{figure}

To test these general predictions, we have first performed Monte Carlo
simulations of the classical FSAFM (see Fig.\ 3 of \cite{prl} for the
magnetization curve at $J'={J \over 2}$). Here, we present results for the
static structure factor $S^{\alpha\beta}(\vec{q}) = {1 \over V^2}
\sum_{\vec{x}} {\rm e}^{i \vec{q} \cdot \vec{x}}
\left\langle S^{\alpha}_{\vec{0}} S^{\beta}_{\vec{x}} \right\rangle$
($V$ is the total number of spins)
on a $24 \times 24$ cluster in Fig.~\ref{SCfig}. The top panel shows
that at $h=4J$ non-vanishing harmonics in the longitudinal structure
factor both at $\vec{q}=(\pi,0)$ and $\vec{q}=(\pi,\pi)$
exist only for $0.494<{J' \over J}<0.508$. Neither, a lower field of $h=2J$,
nor a higher field of $h=6J$ induces a nonzero value of $S^{zz}(\vec{q})$
at these points in the Brillouin zone. The lower panel of
Fig.~\ref{SCfig} presents the transverse structure factor.
In the region of weak (strong) diagonal
coupling $J'$ one identifies N\'eel (striped) types of spin configuration.
At $h=4 J$ the corresponding peaks jump to zero at
exactly the same points where the peaks in the longitudinal
structure factor appear, identifying the spin configuration in
the intermediate region as the `uuud' state. For $h=2 J$ and $h=6 J$,
no significant peak is observed in any of the structure factors for
$J' \approx {J \over 2}$, indicating disordered states above and below
the uuud state.

\begin{figure}
\centering
\includegraphics[angle=270,width=7.5 true cm]{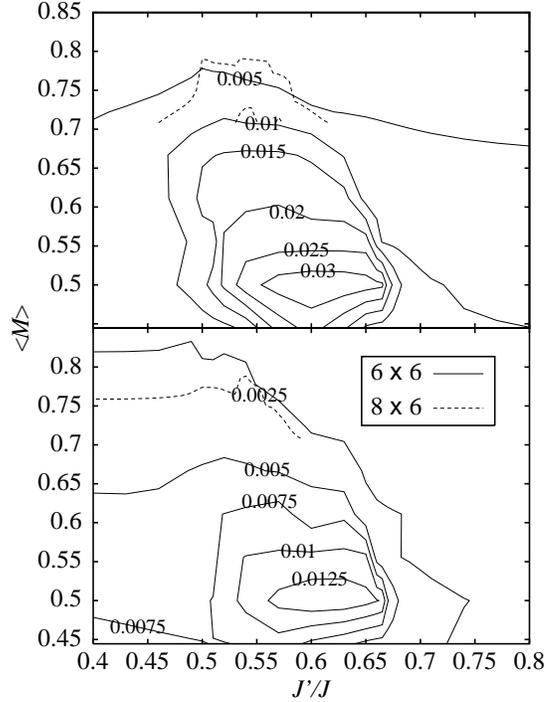}
\caption{Contour plot of $S^{zz}(\pi,0)$ (top panel) and
$S^{zz}(\pi,\pi)$ (bottom panel) for the $S={1 \over 2}$ FSAFM at $T=0$.}
\label{intFig}
\end{figure}

Finally, we study the $S={1 \over 2}$ FSAFM using Lanczos diagonalization.
A detailed discussion \cite{prl,proc} of the magnetization
curves and the static structure factors shows that an
$\langle M \rangle = {1 \over 2}$ plateau exists in the region
$0.5 \lesssim {J' \over J} \lesssim 0.65$ and that it
corresponds to a state with uuud order.
The finite-size analysis of states with $\langle M \rangle \ne {1 \over 2}$
suffers from the problem that data for different system sizes
is available only for very restricted values of $\langle M \rangle$.
Here we propose to interpolate the data for a given cluster size. The
contour-lines which one obtains in the
${J' \over J}$-$\langle M \rangle$-plane are shown in Fig.\ \ref{intFig} for
$6 \times 6$ and $8 \times 6$ clusters. One indeed observes pronounced peaks
on the $6 \times 6$ cluster at $\langle M \rangle = {1 \over 2}$ which
characterize
the uuud state. The comparison with the data for the $8 \times 6$ cluster
(which is available only for $\langle M \rangle \ge {17 \over 24} \approx
0.70833$) suggests that the peak in $S^{zz}(\pi,0)$ survives up to
magnetizations as large as $\langle M \rangle \approx 0.75$ whereas
$S^{zz}(\pi,\pi)$ presumably vanishes in this range. Although it
remains to be clarified whether this (partial) order for
$\langle M \rangle > {1 \over 2}$ in the longitudinal components coexists
with order
in the transverse components (e.g.\ in $S^{xx}(\pi,\pi)$), this signals a
supersolid phase above the uuud state. This scenario is in sharp contrast
to the classical FSAFM, but similar to the 2D $S={1 \over 2}$
Shastry-Sutherland model where supersolid states are presumably
realized above the $\langle M \rangle = {1 \over 3}$ and
$\langle M \rangle = {1 \over 2}$ plateaux \cite{MoTo}.

To summarize, thermal as well as quantum fluctuations
stabilize a collinear uuud state in the FSAFM for $J' \approx {J \over 2}$ at
$\langle M \rangle = {1 \over 2}$.
For $S={1 \over 2}$, this leads to a pronounced plateau in the magnetization
curve. In contrast to these similarities at $\langle M \rangle = {1 \over 2}$,
the classical and quantum model presumably differ for magnetic fields
above this uuud state: Disorder in the classical system versus
a partially ordered, supersolid state in the quantum system. 


%
%
%
%


%
%
%
%



\begin{thebibliography}{00}

\bibitem{Kageyama-Onizuka}
    H.\ Kageyama {\it et al.}, Phys.\ Rev.\ Lett.\ {\bf 82} (1999) 3186;
    K.\ Onizuka {\it et al.}, J.\ Phys.\ Soc.\ Jpn.\ {\bf 69} (2000) 1016.
\bibitem{MoTo} T.\ Momoi, K.\ Totsuka, Phys.\ Rev.\ {\bf B62} (2000) 15067.
\bibitem{prl} M.E.\ Zhitomirsky, A.\ Honecker, O.A.\ Petrenko,
              Phys.\ Rev.\ Lett.\ {\bf 85} (2000) 3269.
\bibitem{proc} A.\ Honecker, cond-mat/0009006, to appear in Can.\ J.\ Phys.
\bibitem{MCLMTMM} R.\ Melzi {\it et al.}, Phys.\ Rev.\ Lett.\ {\bf 85} (2000)
              1318.
\bibitem{mzh} M.E.\ Zhitomirsky, cond-mat/0107608.
\bibitem{Wal-privcom}
              Ch.\ Waldtmann {\it et al.}, Eur.\ Phys.\ J.\ {\bf B2} (1998) 501;
              Ch.\ Waldtmann, H.-U.\ Everts, private communication.

\end{thebibliography}
\end{document}